\documentclass[aps,amssymb,twocolumn,superscriptaddress,nofootinbib]{revtex4}
\usepackage{setspace}
\usepackage{graphicx}
\usepackage{psfrag}

\newcommand{\Screll}{{\mathcal L}}

\def\be{\begin{equation}}
\def\ee{\end{equation}  }
\def\bea{\begin{eqnarray}}
\def\eea{\end{eqnarray}  }
\def\rg{\sqrt{-g}}

\begin{document}
\title{Numerical Relativity Using a Generalized Harmonic Decomposition}

\author{Frans Pretorius}
\affiliation{Theoretical Astrophysics,
     California Institute of Technology,
     Pasadena, CA 91125}

\begin{abstract}
A new numerical scheme to solve the Einstein field equations
based upon the generalized harmonic decomposition of the Ricci tensor
is introduced.
The source functions driving the wave equations that define generalized
harmonic coordinates are treated as independent functions, and encode
the coordinate freedom of solutions. Techniques are discussed to impose
particular gauge conditions through a specification of the source functions.
A 3D, free evolution, finite difference code implementing this system of 
equations with a scalar field matter source is described. The 
second-order-in-space-and-time 
partial differential equations are discretized directly
without the use first order auxiliary terms,
limiting the number of independent 
functions to fifteen---ten metric quantities, four source functions and the
scalar field. This also limits the number of constraint equations, which can only
be enforced to within truncation error in a numerical free evolution, to four.
The coordinate system is compactified to spatial infinity in order to
impose physically motivated, constraint-preserving outer boundary conditions.
A variant of the Cartoon method for efficiently simulating axisymmetric 
spacetimes with a Cartesian code is described that does not use interpolation,
and is easier to incorporate into existing adaptive mesh refinement packages.
Preliminary test simulations of vacuum black hole evolution and black hole
formation via scalar field collapse are described, suggesting that this method 
may be useful for studying many spacetimes of interest.

\end{abstract}

\maketitle

\section{Introduction}

One of the primary goals of numerical relativity today is to solve for 
astrophysical spacetimes that are expected to be strong sources of gravitational
wave emission in the frequency bands relevant to current and planned gravitational
wave detectors. Expected sources include the inspiral and merger of compact objects,
supernovae, pulsars, and the big bang. An important
tool for extracting physics from detector signals is the technique of matched filtering,
which requires an accurate wave form of a model of the expected source. 
For binary black hole mergers (in particular) it is thought that numerical relativity
is the only method that will be able to provide such waveforms close to and during
the plunge phase of the merger. Despite significant progress made over the past decade,
a full solution to this problem still eludes researches. One reason for the difficulty is
the complexity of the field equations. This translates into significant computer resources
being needed to solve the equations, which limits the turn-around time for testing new ideas.
However, perhaps the largest obstacle so far has been finding a formalism to write the
field equations in that is amenable to long-term, stable numerical evolution. Some of 
the promising techniques used today include symmetric hyperbolic 
formalisms \cite{reula_review,lehner_review},
the BSSN formalism (sometimes referred to as the NOK formalism) \cite{nok,sn,bs,bruegmann_et_al} 
and characteristic 
evolution (for black hole/neutron star systems) \cite{bishop_et_al}. 
Several groups are also beginning
to examine the possibility of constrained evolution for the 3D binary black hole 
problem \cite{anderson_matzner,bonazzola_et_al,spectral_group}, and other
promising directions make use of tetrad formulations 
of the field equations \cite{erw,bb,estabrook}, and solution of the
conformal field equations \cite{friedrich_c1,friedrich_c2,hubner,husa}.

A method of writing the field equations that has proven very useful
in analytic studies is arrived at by imposing the 
{\em harmonic coordinate condition}, where the four spacetime coordinates
$x^\mu$ are chosen to individually satisfy wave equations: $\Box x^\mu=0$.
The Einstein equations, when written with this condition imposed, take on
a mathematically appealing form where the principal part of each partial
differential equation satisfied by a metric component $g_{\alpha\beta}$
becomes the scalar wave operator $\Box g_{\alpha\beta}$. This allowed 
for (among other things) the first existence and uniqueness proof
of solutions to the field equations \cite{bruhat}. In numerical relativity,
a solution scheme based directly upon this formulation of the field
equations has recently been suggested by Garfinkle \cite{garfinkle} (see also
related work by Szilagyi and Winicour \cite{szilagyi_winicour},
and the so-called Z4 system\cite{z4}, which seems to be quite similar
to generalized harmonic evolution in many respects). Garfinkle
considered a generalization of the harmonic coordinate condition
of the form $\Box x^\mu=H_\mu$, where $H_\mu$ are now arbitrary source 
functions, and found that the technique was successful in simulations
of the approach to the singularity in certain cosmological spacetimes.

One purpose of this paper is to begin to investigate the use of the generalized harmonic
decomposition in asymptotically flat spacetimes. 
The formalism is described in Sec. \ref{sec_formalism}. If this
method is to be useful for a large class of spacetimes, one issue
that needs to be addressed is how to choose gauge conditions via
specification of the source functions $H_\mu$; this topic is discussed in 
Sec. \ref{sec_gauge}. A second goal of this paper is to investigate 
direct discretization of the second-order-in-space-and-time
partial differential equations\footnote{Recent analytic 
investigations by Calabrese\cite{calabrese_pc} have
suggested that such a scheme may suffer from high-frequency instabilities
in situations where the coefficients in front of mixed time-space derivatives
are greater than the local characteristic speed. We have not yet noticed such an instability,
probably because of the numerical dissipation we use, which was one of the
suggested cures for the problem in \cite{calabrese_pc}.}
(in other
words, the system is {\em not} converted to a system of first order equations
before discretization).
One reason for doing so is to have a free evolution scheme where the {\em only}
constraints amongst the variables are the four constraint equations imposed by the
Einstein equations (see also \cite{kreiss_ortiz,szilagyi_winicour}). 
The hope then is that even if this system suffers from 
``constraint violating modes'' \footnote{By constraint violating mode we mean a
solution to the {\em continuum evolution} equations that is not a
solution of the full Einstein equations, and furthermore exhibits exponential
growth from initial data with arbitrarily small deviations from putative
initial data that does satisfy the constraints.}, it may 
be easier to analyze and cure them using (for instance) ideas suggested in research of
symmetric hyperbolic versions of the field equations \cite{kelly_et_al,shinkai_et_al,
tiglio,tiglio_et_al,lindblom_et_al}\footnote{For it appears that it
may {\em not} be possible to construct a constrained-transport type numerical 
evolution scheme that satisfies
all of the Einstein equations to machine precision \cite{meier,bartolo_et_al}.}. The numerical
code is described in Sec. \ref{sec_code}, along with related topics such
as apparent horizon finding, excision, boundary conditions, initial conditions, and
the current scalar field matter source. Also described in Sec. \ref{sec_code}
is a variant of the {\em Cartoon} method \cite{cartoon} to efficiently simulate axisymmetric 
spacetimes with a Cartesian code. The advantages of the method presented here
are that no interpolation is used, and the axisymmetric simulation is performed
on a two-dimensional slice of the Cartesian grid. In Sec. \ref{sec_results} 
test simulations of black hole evolution and gravitational collapse are shown,
suggesting that this solution method holds promise for simulating asymptotically
flat spacetimes. Concluding remarks are given in Sec. \ref{sec_conclusion}, in particular
a discussion of some of 
the work that still needs to be done before the code could provide new physical
results in situations of interest.

\section{The Einstein Field Equations in the Generalized Harmonic Decomposition}\label{sec_formalism}

Consider the Einstein field equations in the form
\be\label{efe}
R_{\alpha\beta}=4\pi\left(2 T_{\alpha\beta}-g_{\alpha\beta} T\right),
\ee
where $R_{\alpha\beta}$ is the Ricci tensor, $g_{\alpha\beta}$ is the
metric tensor, $T_{\alpha\beta}$ is the stress energy tensor
with trace $T$, and units have been chosen so that Newton's constant $G$
and the speed of light $c$ are equal to 1. 
The Ricci tensor is defined in terms of the Christoffel
symbols $\Gamma^{\gamma}_{\alpha\beta}$ by
\be\label{ricci}
R_{\alpha\beta}=\Gamma^{\delta}_{\alpha\beta,\delta} - 
                \Gamma^{\delta}_{\delta\beta,\alpha}
               +\Gamma^{\epsilon}_{\alpha\beta}\Gamma^{\delta}_{\epsilon\delta}
               -\Gamma^{\epsilon}_{\delta\beta}\Gamma^{\delta}_{\epsilon\alpha}
\ee
where $\Gamma^{\gamma}_{\alpha\beta}$ is
\be
\Gamma^{\gamma}_{\alpha\beta}=\frac{1}{2}
g^{\gamma\epsilon}\left[g_{\alpha\epsilon,\beta}+g_{\beta\epsilon,\alpha}-g_{\alpha\beta,\epsilon}
\right]
\ee
The notation $f_{,\alpha}$ and $\partial_{\alpha}f$ is used interchangeably to
denote ordinary differentiation of some quantity $f$ with respect to the coordinate
$x^\alpha$. 

Introduce a set of four {\em source functions} $H^\mu$ via
\bea
H^\mu &\equiv& \Box x^\mu \\
      &=& \frac{1}{\rg}\partial_\alpha\left(\rg g^{\alpha\beta} x^\mu_{,\beta}\right) \\
      &=& \frac{1}{\rg}\partial_\alpha\left(\rg g^{\alpha\mu} \right),
\eea
or equivalently, defining $H_\mu = g_{\mu\nu} H^\nu$, we have
\be\label{hdef}
H_\mu = \left(\ln\rg\right)_{,\mu} - g^{\alpha\nu}g_{\nu\mu,\alpha}.
\ee
The symmetrized gradient of $H_\mu$ is thus
\be\label{dhdef}
H_{(\mu,\nu)} = (\ln\rg)_{,\mu\nu} - g^{\alpha\beta}{}_{(,\nu}g_{\mu)\beta,\alpha}
                                 - g^{\alpha\beta}g_{\beta(\mu,\nu)\alpha}
\ee
The generalized harmonic decomposition involves replacing particular
combinations of first and second derivatives of the metric in the Ricci 
tensor (\ref{ricci}) by the equivalent quantities in 
(\ref{hdef},\ref{dhdef}), and then {\em promoting the source functions $H_{\mu}$
to the status of independent quantities}. 
Specifically, one can rewrite the field equations (\ref{efe}) as
\bea
g^{\delta\gamma}g_{\alpha\beta,\gamma\delta} 
+ g^{\gamma\delta}{}_{,\beta} g_{\alpha\delta,\gamma}
+ g^{\gamma\delta}{}_{,\alpha} g_{\beta\delta,\gamma}  
+ 2 H_{(\alpha,\beta)} \nonumber\\
- 2 H_\delta \Gamma^\delta_{\alpha\beta} 
+2 \Gamma^\gamma_{\delta\beta}\Gamma^\delta_{\gamma\alpha} 
= - 8\pi\left(2 T_{\alpha\beta}-g_{\alpha\beta} \label{efe_h}T\right)
\eea
As $H_{\mu}$ are now four independent functions,
one needs to provide four additional, independent differential equations to 
solve for them, which we write schematically as
\be\label{he}
\Screll_\mu H_\mu = 0 \ \ \ \mbox{(no summation)}.
\ee
$\Screll_\mu$ is a differential operator that in general can dependent
upon the spacetime coordinates, the metric and its derivatives, and the 
source functions and their derivatives. 
Note however that the principal part of (\ref{efe_h}) 
is now the simple wave operator $g^{\delta\gamma}\partial_\gamma\partial_\delta$
acting upon each metric component $g_{\alpha\beta}$; this subsystem
of equations is manifestly hyperbolic given certain reasonable conditions
on the metric\footnote{For example one would need a single coordinate to be
timelike and the rest to be spacelike throughout the integration volume.} 
and as long as the coupling between (\ref{efe_h}),
(\ref{he}) and any matter evolution equations that may be needed
do not the affect the characteristic structure of (\ref{efe_h}). We will
not discuss the well-posedness of this system of equations here, though
this is certainly a topic worth pursuing.

The Einstein field equations are thus equivalent to the system of equations
(\ref{efe_h}) and (\ref{he}), {\em provided} that the harmonic ``constraints'' (\ref{hdef}) 
are satisfied for all time $t\equiv x^0$. The claim then is, at the analytical 
level, if (\ref{efe_h}) is
used to evolve $g_{\alpha\beta}$, and (\ref{he}) is used to evolve
$H_\mu$, then (\ref{hdef}) will be satisfied for all time
provided that initial conditions are specified so that (\ref{hdef}) 
{\em and} (\ref{dhdef}) are satisfied then. For the special case
where $H^\mu$ are given as {\em a-priori} functions of the coordinates
$x^\mu$, the preceding statement has been proven before 
\cite{friedrich} (the case $H^\mu=0$ was first shown in \cite{bruhat}).
The idea behind the proof is as follows. Define the harmonic constraint
function $C^\mu$ as 
\be
C^\mu \equiv H^\mu - \Box x^\mu.
\ee
For any solution to the Einstein equations (\ref{efe}), $C^\mu$ is
identically zero. Using the contracted Bianchi identity and 
conservation of stress energy, one can show that $C^\mu$ satisfies
the following homogeneous wave equation
\be\label{h_const}
\Box C^\mu = - R^\mu{}_\nu C^\nu.
\ee
Therefore, given any $g_{\mu\nu}$ that satisfies (\ref{efe_h}) for all time together
with some $H^\mu$ that satisfies both $C^\mu=0$ and $\partial_t C^\mu=0$ at $t=0$,
(\ref{h_const}) guarantees that $g_{\mu\nu}$ will also solve the Einstein 
equations (\ref{efe}) for all time. We cannot prove such a result for a
general evolution system where $H^\mu$ is specified via some arbitrary
set of differential equations. Rather, we will take the 
more pragmatic approach in the numerical code of demonstrating
convergence to a solution of the Einstein equations for any
particular evolution system we use. In fact, such a convergence test
is the {\em only} measure of the validity of the numerical solution,
regardless of any analytic properties of the underlying continuum
problem.

Equivalent to enforcing
(\ref{dhdef}) at $t=0$ is to make sure that the usual constraint 
equations, namely
\bea
^{(3)}R + K^2 - K_{ab} K^{ab} = 16\pi\rho, \label{hc} \\
K_{a\ \ |b}^{\ b} - K_{|a} = 8\pi J_a \label{mc}
\eea
are satisfied then, which from a practical standpoint is easier
to solve than (\ref{dhdef}) using existing, well-established techniques\cite{cook,lehner_review}. 
In the above, $K_{ab}$ is
the extrinsic curvature tensor of the $t={\rm const.}$ hypersurface
with induced metric $h_{ab}$, $K$ is the trace of $K_{ab}$, 
$^{(3)}R$ is the Ricci scalar of $h_{ab}$, $|$ denotes the covariant
derivative operator compatible with $h_{ab}$, and $\rho$ and $J_a$
are the projected matter energy and momentum densities respectively.
Note that we use notation where Greek indices denote four dimensional
quantities and run from $0$ to $3$, and
Latin indices denote three dimensional spatial quantities and run from
$1$ to $3$.

At this stage the system (\ref{efe_h},\ref{he})
is completely general in that we have not yet specified 
any time slicing or spatial coordinates. Choosing a gauge amounts
to specifying a set of source functions through ($\ref{he}$),
and thus the source functions play a role analogous to the
lapse function and shift vector in the traditional ADM decomposition.
One disadvantage of the harmonic decomposition is that (to my knowledge)
there is no simple geometric description of the relationship
between $H_\mu$ and the resulting spacetime coordinates. In same cases
one can appeal to the ADM lapse and shift view of coordinate freedom 
to motivate a particular choice of $H_\mu$. We will discuss these
and several other classes of gauge conditions that 
may be useful for numerical evolution in the following section.

\section{Specifying a Gauge}\label{sec_gauge}
Within the generalized harmonic decomposition one can think
of the source functions $H_\mu$ as representing the four
coordinate degrees of freedom available in general relativity.
There are many conceivable ways of choosing $H_\mu$; in this
section we will give a few suggestions, several
of which are used in the evolutions presented in Sec. \ref{sec_results}.
However, the discussion here is rather heuristic in that we
do not consider how any of these gauge choices may affect the
character of the coupled Einstein-gauge evolution system.
Note that gauge source functions were discussed by Friedrich\cite{friedrich2}
in some detail, though not specifically within the context
of supplying additional evolution equations for them.

The simplest gauge choice in this formalism is to set the
source functions equal to some arbitrary functions of the
spacetime coordinates:
\be\label{harm_gauge}
H_\mu=f_\mu(x^\alpha).
\ee
The case $f_\mu=0$ is standard harmonic coordinates.
The next condition we consider is a 
coordinate system that evolves toward harmonic coordinates:
\be\label{harm_dgauge}
\partial_t H_\mu= - \kappa_\mu(t) H_\mu \ \ \ \mbox{(no summation)},
\ee
where $\kappa_\mu$ are a set of 4 arbitrary though positive functions of time,
which if non-zero will cause $H_\mu$ to evolve to zero.

A useful method to derive coordinate conditions for the harmonic
decomposition is to appeal to the manner in which the coordinate
system is specified in the ADM decomposition. This makes available
a tremendous amount of research that has gone into gauge related issues
for ADM-based evolution \cite{lehner_review}. In the ADM formalism
the metric element is written as
\be\label{adm}
ds^2 = -\alpha^2 + h_{ij} \left(dx^i + \beta^idt\right)\left(dx^j + \beta^jdt\right),
\ee
where the {\em lapse function}  $\alpha$ measures the rate of change of
proper time with respect to coordinate time $t$ of hypersurface normal observers,
$h_{ij}$ is the intrinsic metric of $t=\rm{const.}$ slices, and 
the {\em shift vector} $\beta^j$ describes how the spatial coordinates change
for normal observers from one time slice to the next. The normal component
and spatial projection of the source function $H_\mu$ are \footnote{Note however
that $H_\mu$ does not transform like a one-form under coordinate transformations,
and hence the projections are not covariant objects.}
\bea
H\cdot n &\equiv& H_\mu n^\mu \nonumber\\
         &=& -K - \partial_{\nu} (\ln\alpha) n^\nu \label{hdotn} \\
\perp H^i &\equiv& H_\mu h^{\mu i} \nonumber\\
          &=& - \bar{\Gamma}^i_{jk} h^{jk}
  + \partial_j (\ln\alpha) h^{i j} + \frac{1}{\alpha}\partial_\gamma \beta^i n^\gamma \label{hdoth},
\eea
where $\bar{\Gamma}^i_{jk}$ is the connection associated with $h_{ij}$,
and $n^\nu$ is the hypersurface normal vector given by
\be\label{norm}
n_\nu = -\alpha \partial_\nu t
\ee
Notice that in (\ref{hdotn}) and (\ref{hdoth}) the time derivative of $\alpha$ only appears 
in $H\cdot n$, while the time derivative of $\beta^i$ only appears in the corresponding
component of $\perp H^i$. In other words, the choice of the normal component $H\cdot n$
in an evolution directly affects the rate of change of $\alpha$ with respect to time,
and therefore $H\cdot n$ controls the time-slicing of the spacetime; similarly, $\perp H^i$
controls the manner in which the spatial coordinates evolve with time
(another way of stating this is that (\ref{hdotn},\ref{hdoth}) are generalizations of the 
hyperbolic equations governing the lapse and shift
within harmonic coordinates\cite{york}).
One way in which an ADM style gauge condition can be used within
the harmonic decomposition is to substitute the corresponding choices
of $\alpha$ and $\beta^i$ into (\ref{hdotn}-\ref{hdoth}), and use
the result as the source functions for the harmonic evolution. 
The simplest class of gauge conditions that can be implement in this fashion
are the so called ``driver'' conditions 
\cite{bona_masso,balakrishna_et_al,alcubierre_et_al,lindblom_scheel,alcubierre_et_al_2}, 
where one directly specifies the time derivatives of $\alpha$ and $\beta^i$
to achieve, for example, approximate maximal slicing and minimal
distortion gauges respectively.

The manner in which the ADM driver conditions are implemented 
suggests a similar way in which such gauge conditions can
be used in a harmonic evolution: instead of substituting in
the forms for $\alpha$ and $\beta^i$ in (\ref{hdotn}-\ref{hdoth}) to
try to satisfy the conditions exactly, choose source functions
to {\em drive} the gauge toward the desired one. To see how this can be done,
first rewrite (\ref{hdotn}-\ref{hdoth}) as evolution equations for the 
the lapse and shift:
\bea
\partial_t \alpha = - \alpha^2 H\cdot n + ... \label{ldot} \\
\partial_t \beta^i = \alpha^2 \perp H^i + ... \label{bdot},
\eea
where the ellipses denote the rest of the terms that do not
contain $\partial_t \alpha,\partial_t \beta^i$ or $H^\mu$.
Now suppose at some instant of time the {\em desired} value
of the lapse and shift are calculated (by whatever means)
to be $\alpha_0$ and $\beta^i_0$ respectively. Then from
(\ref{ldot}-\ref{bdot}) {\em one possible}
set of choices for the source functions that will cause the lapse
and shift to evolve toward the desired values are
\bea
H\cdot n &=& \kappa_n(t) \frac{\alpha-\alpha_0}{\alpha^2} \label{hdotn_b}\\
\perp H^i &=& - \kappa_i(t) \frac{\beta^i-\beta^i_0}{\alpha^2} \label{perph} \ \ \ \mbox{(no summation)}, 
\eea
where $\kappa_n$ and $\kappa_i$ are positive functions of time that
can be used to control the rate of evolution. 

It many circumstances it may make more sense to implement the above
style driver conditions as evolution equations, rather than algebraic
conditions. This could be, for example, if the initial conditions for
$H_\mu$ are not compatible with the desired gauge choice, and
so implementing (\ref{hdotn_b}-\ref{perph}) will result in discontinuous
source functions at the initial time. A couple of alternative possibilities include
\be
\frac{\partial H\cdot n}{\partial t} = \kappa_n(t) \frac{\alpha-\alpha_0}{\alpha^2} \label{dot_hdotn}\\
\ee
and
\be
\Box(H\cdot n) = - \kappa_n(t) \frac{\alpha-\alpha_0}{\alpha^2} + 
                    \xi_n(t) (H\cdot n)_{,\mu} n^\mu \label{box_hdotn},
\ee
with similar expressions for the spatial parts of $H_\mu$. $\xi_n(t)$ is 
a positive function that can be used to add a dissipative term to ($\ref{box_hdotn}$).
One advantage of using a wave operator ($\ref{box_hdotn}$) to evolve the source functions
is then the principal parts of all equations in the system (\ref{efe_h}-\ref{he}) have the same 
characteristic structure
(as long as $\alpha_0$ and $\beta^i_0$ depend
at most on first derivatives of the fundamental variables). This may be important to establish
well-posedness of the coupled system of equations \cite{cbona}.

It is beyond the scope of this paper to investigate how well any of these
suggested gauge conditions perform in situations of interest, however in 
Sec. \ref{sec_results} we will show some preliminary results indicating
that these ideas can be implemented in a stable fashion.

\section{Numerical Code}\label{sec_code}
In this section we describe a 3D numerical code based upon the generalized harmonic
decomposition. This code has several features of note
\begin{itemize}
\item {\em Direct discretization of (\ref{efe_h}-\ref{he})}: In other words, we do not
convert the system of equations to first order form---the only 
variables used are the 10 unique metric components $g_{\alpha\beta}$, 4 source
functions $H_\mu$, and matter variables.
\item {\em Spatially compactified Cartesian coordinates}: The spatial coordinates
are compactified to include $i^0$, to simplify the imposition of physically
realistic boundary conditions for asymptotically flat space times.
\item {\em Black hole excision}: Black hole excision is used to evolve spacetimes
containing black holes, whereby portions of the computational domain
inside of apparent horizons are ``excised'' to remove the singularities.
\item {\em Built within a parallel adaptive mesh refinement (AMR) framework}:
The code utilizes a new set of parallel AMR libraries which we will describe elsewhere,
though the Berger and Oliger style AMR algorithm used is very similar to the one 
presented in \cite{fp_mgamr,fp_thesis}.
\item {\em Efficient simulation of axisymmetric spacetimes using a variant of the
Cartoon method \cite{cartoon}}: The algorithm presented here does not require
interpolation, and only utilizes a single 2-dimensional slice of the
Cartesian grid, simplifying incorporation into existing AMR packages.

\end{itemize} 
In the remainder of this section we will describe certain aspects of the code in more detail.

\subsection{Discretization Scheme}\label{sec_disc}

From here onward we will use the coordinate names $t\equiv x^0$ and
$(x,y,z)\equiv(x^1,x^2,x^3)$. Also, as discussed in the next section,
we use a compactified coordinate system in the code. This necessitates 
the use of regularized variables for some of the metric and source function
components, however to keep the discussion in this section simpler
we ignore that aspect of the code here.
In Appendix \ref{app_stab} we present a stability analysis of this
discretization method applied to a one-dimensional wave equation in
flat space. The purpose of the analysis is to give a simple, concrete
example of the numerical method, and to show that there are no 
fundamental instabilities in it. Of course, this cannot prove that
the full, non-linear problem in compactified coordinates will be 
stable---doing so is beyond the scope of this paper.

\begin{figure}
\begin{center}
\psfrag{fnp1i}{$f^{n+1}_{i}$}
\psfrag{fni}{$f^{n}_{i}$}
\psfrag{fnip1}{$f^{n}_{i+1}$}
\psfrag{fnim1}{$f^{n}_{i-1}$}
\psfrag{fnm1i}{$f^{n-1}_{i}$}
\psfrag{tn}{$t^n$}
\psfrag{tnp1}{$t^{n+1}$}
\psfrag{tnm1}{$t^{n-1}$}
\psfrag{xi}{$x_i$}
\psfrag{xip1}{$x_{i+1}$}
\psfrag{xim1}{$x_{i-1}$}
\includegraphics[width=7.0cm,clip=true]{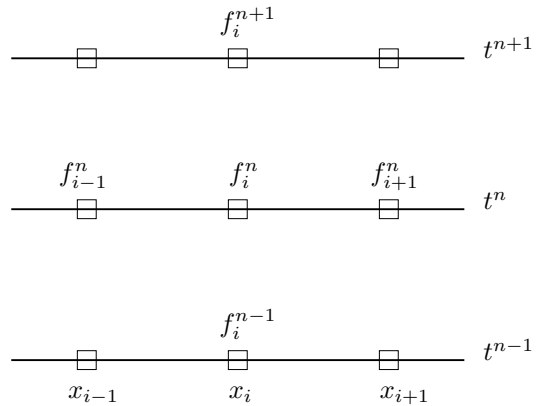}
\end{center}
\caption {The discretization of a variable $f(t,x,y,z)$ in the $x-t$ plane.
}
\label{sample_mesh}
\end{figure}

We use second order accurate finite difference techniques to discretized
(\ref{efe_h}-\ref{he}) and the scalar field evolution equation (\ref{phi_eom}) presented
in Sec. \ref{sec_matter}. This is a set of 15 equations for 15 unknown functions---the 10
non-trivial metric components $g_{\alpha\beta}$, the 4 source functions $H_\mu$
and the scalar field $\Phi$. In the discretized version
of (\ref{efe_h}) all Christoffel symbols, contravariant metric elements and their
gradients are replaced with the appropriate sum of covariant metric elements and
their gradients. As (\ref{efe_h}) and (\ref{phi_eom}) are second order partial differential
equations in time, second order accurate discretization requires a 3 time level scheme 
(at a minimum). Fig. \ref{sample_mesh} shows a schematic representation 
(with two spatial dimensions suppressed) of the discretization of a variable $f(t,x,y,z)$.
$f(t,x,y,z)$ evaluated at a grid location 
$(t^n,x_i,y_j,z_k)$ = $(n \Delta t,i \Delta x,j \Delta y,k \Delta z)$
is denoted by $f^n_{ijk}$, where $n,i,j$ and $k$ are integers, and $\Delta t$, $\Delta x$, 
$\Delta y$ and $\Delta z$ are the temporal and spatial discretization scales respectively.
Table \ref{ops} contains a representative sample of the finite difference operators used to
evaluate derivatives on the mesh. Replacing the continuum variables with discrete variables,
and the derivative operators with difference operators will result in a difference
equation 
\be\label{fde}
\Screll_f|^n_{ijk} = 0
\ee
for each variable $f$ at each grid point $(t,x,y,z)$=$(t^n,x_i,y_j,z_k)$ 
in the computational domain. 

We solve the system of equations (\ref{fde}) using a Newton-Gauss-Seidel relaxation 
scheme, as follows.
Initial data for a single time step at $t=t^n$ consists of all the 
variables at time levels $t^n$ and $t^{n-1}$. The unknowns are the variables at 
time level $t^{n+1}$. Denote an approximate value of the unknown $f^{n+1}_{ijk}$
by $\hat{f}^{n+1}_{ijk}$. The iteration is set up using function values
at time level $n$ as an initial guess to the solution at time level $n+1$. One step
of the iteration then proceeds by updating each unknown, in turn, via
\be\label{ngs}
\hat{f}^{n+1}_{ijk} \rightarrow \hat{f}^{n+1}_{ijk} - \frac{\mathcal{R}_f|^n_{ijk}}{\mathcal{J}_f|^n_{ijk}},
\ee
where $\mathcal{R}_f|^n_{ijk}$ is the residual of the difference equation (the left hand
side of (\ref{fde}) evaluated using the approximate solution) and $\mathcal{J}_f|^n_{ijk}$
is the diagonal element of its Jacobian
\be
\mathcal{J}_f|^n_{ijk} = \frac{\partial \Screll_f|^n_{ijk}}{\partial f^{n+1}_{ijk}},
\ee
again evaluated with the approximate solution. In other words, (\ref{ngs}) is simply
solving a linearized version of (\ref{fde}) for $f^{n+1}_{ijk}$ assuming all other
unknowns are fixed. The iteration is repeated until the residual for all variables
is below some specified tolerance.

\begin{table}
\begin{tabular}[t]{| c || c |}
\hline
$f_{,x}$  & $(f^n_{i+1}-f^n_{i-1})/(2\Delta x)$ \\
          & \\
$f_{,t}$  & $(f^{n+1}_i-f^{n-1}_i)/(2\Delta t)$ \\
          & \\
$f_{,xx}$ & $(f^n_{i+1}-2f^n_i+f^n_{i-1})/(\Delta x)^2$ \\
          & \\
$f_{,tt}$ & $(f^{n+1}_i-2f^n_i+f^{n-1}_i)/(\Delta t)^2$ \\
          & \\
$f_{,tx}$ & $(f^{n+1}_{i+1}-f^{n+1}_{i-1}-f^{n-1}_{i+1}+f^{n-1}_{i-1})/(4\Delta x\Delta t)$ \\
\hline
\end{tabular}
\caption{A sample of the finite difference stencils used to convert
the differential equations to difference equations. The column
on the right shows the second order accurate representation (with $y$ and
$z$ indices suppressed for clarity) of the corresponding derivative
operator to the left, evaluated at the point $(t^n,x_i,y_j,z_k)$.
Similar stencils are used for terms containing $y$ and $z$ derivatives.
}
\label{ops}
\end{table}

\subsubsection{Numerical Dissipation}\label{sec_diss}

Some form of numerical dissipation is necessary to stably evolve certain
spacetimes, in particular those that contain black holes. We use Kreiss-Oliger 
style dissipation \cite{KO}, however, rather than modify the discrete
evolution equations as is typically done (and note also that \cite{KO} considered
first order in time systems), we apply the dissipation 
as a {\em filter} to the discrete variables, at {\em both} past time levels
$t^n$ and $t^{n-1}$, {\em prior} to updating $t^{n+1}$. 
Specifically,
at a given time level we define the high-frequency component $\eta^x_{ijk}$ 
of grid function $f_{ijk}$, in the $x$ direction as 
\bea
\eta^x_{ijk} &=& \frac{1}{16}(f_{i-2jk}-4f_{i-1jk}+6f_{ijk} \nonumber \\ 
             &\ &\ \ \ \ \ -4f_{i+1jk}+f_{i+2jk}), \ \ \ 2<i<N_x-2 \nonumber \\
             &=& 0 \ \ \ {\rm elsewhere}, \label{x_hf}
\eea
where the local size of the mesh is $N_x$ points in the $x$ direction.
After $\eta^x_{ijk}$ has been calculated over the entire local grid, 
it is subtracted from $f$ as follows
\be
f_{ijk} = f_{ijk} - \epsilon \eta^x_{ijk}, \label{x_hf_sub}
\ee
where $\epsilon$ is a constant, required to be in the range $0..1$ for
stability. In practice we use values of $\epsilon$ in the range $0.2$ to $0.5$.
Once the high frequency components in the $x$ direction have been 
subtracted, the procedure is repeated for the high frequency components
$\eta^y_{ijk}$ and $\eta^z_{ijk}$ of $f_{ijk}$ in the $y$ and
$z$ directions respectively, which are given by expressions similar
to (\ref{x_hf}).

We did experiment with extending the dissipation filter to 
the grid boundaries as outlined in \cite{calabrese_et_al},
however this did not seem to have a significant effect
on the solution in most circumstances, and seemed to produce
more error (as measured by residuals of the field equations) 
next to excision boundaries without offering improved stability.
However, the excision method proposed in \cite{calabrese_et_al}
was for cubical excision boundaries, and for schemes satisfying
summation by parts, so it is questionable how appropriate it
is to apply that method here.
Also note that applying the above filters to {\em both} past time levels
at each evolution step is essential for long term stability. We do
not know why this is so important; naively one would think
that only applying the filter to time level $t^n$ would be sufficient,
as the update step does not alter the variables at $t^n$,
and since $t^n$ is copied to $t^{n-1}$ after each update step, both
$t^n$ and $t^{n-1}$ are effectively smoothed. Also, a simple
extension of the analysis in Appendix \ref{app_stab} to account
for different amounts of dissipation applied to each of the
past time levels shows that the one dimensional, flat space
wave equation remains stable; hence the need to dissipate both
time levels is particular to black hole spacetimes as
far as we can tell.

\subsection{Coordinate System and Boundary Conditions}\label{sec_coords}

To simplify the imposition of asymptotically flat boundary conditions
we use the following spatially compactified coordinate system.
First, consider an uncompactified Cartesian coordinate system
of the form
\be\label{ucm}
ds^2 = \bar{g}_{tt} dt^2 + 2 \bar{g}_{ti} d\bar{x}^i dt + 
                    \bar{g}_{ij} d\bar{x}^i d\bar{x}^j
\ee
Here $(\bar{x}^1,\bar{x}^2,\bar{x}^3)\equiv(\bar{x},\bar{y},\bar{z})$ 
runs from $-\infty$ to $+\infty$, and in the
limit where $\bar{x}^i \rightarrow \pm\infty$ the metric becomes
the Minkowski metric
\be
ds^2 = -dt^2 + d\bar{x}^2 + d\bar{y}^2 +d\bar{z}^2.
\ee
The following coordinate transformation
\be\label{ct}
\bar{x}^i=\tan(\pi x^i/2)
\ee
(with $\bar{t}=t$) will bring (\ref{ucm}) into the form
\be\label{cm}
ds^2 = g_{tt} dt^2 + 2 g_{ti} dx^i dt + 
                     g_{ij} dx^i dx^j,
\ee
where
\bea
g_{ti} = \frac{\pi}{2} \sec^2(\pi x^i/2) \bar{g}_{ti}, \nonumber \\
g_{ij} = \frac{\pi^2}{4} \sec^2(\pi x^i/2) \sec^2(\pi x^j/2) \bar{g}_{ij} \label{reg_g_def}.
\eea
Now $x^i$ runs from $-1$ to $1$, and spacelike infinity $i^0$ corresponds
to the surfaces $x^i=\pm 1$. Note that in this limit the compactified (unbarred) metric
elements are singular, however the uncompactified parts are still well behaved and
asymptote to their Minkowski values. In the code we thus evolve the
uncompactified components $g_{tt},\bar{g}_{ti}$ and $\bar{g}_{ij}$, analytically 
substituting the values (\ref{reg_g_def}) into (\ref{efe_h}) prior to discretization. 
Furthermore, in the compactified coordinate system (\ref{cm}) 
we {\em define} the spatial source functions
$H_i$ to take the form
\be\label{barhdef}
H_i = \bar{H}_i - \pi \tan(\pi x^i/2).\
\ee
and evolve only the regularized components $\bar{H}_i$ (for note that
in compactified Minkowski coordinates $H_i = \pi \tan(\pi x^i/2)$ 
from (\ref{hdef})).
Therefore, the outer boundary conditions we impose on the regularized
metric and source functions are
\bea
\bar{g}_{tt}(t,i^0) &=& -1 \nonumber \\
\bar{g}_{ti}(t,i^0) &=& 0 \nonumber \\
\bar{g}_{ii}(t,i^0) &=& 1 \nonumber \\
\bar{g}_{ij}(t,i^0) &=& 0, \ \ i\ne j \nonumber \\
H_{t}(t,i^0) &=& 0, \nonumber \\
\bar{H}_i(t,i^0) &=& 0,
\eea
where the notation $(t,i^0)$ refers to any one of the six boundaries
$x=\pm 1$, $y=\pm 1$ and $z=\pm 1$.

We conclude this section by discussing several concerns
about evolving the field equations in a coordinate system compactified
to spatial infinity. It is beyond the scope of this paper to analyze
these issues in more detail, however, we are currently investigating
them. However, note that a similar compactification scheme was used
to model black strings in 5 dimensions \cite{black_string}, with no
notable adverse effects. 

First, the metric, hence equations, are formally singular at $i^0$. The
singular behavior is dealt with using regularized variables and enforcing
Minkowski space boundary conditions at $i^0$, as described above. Nevertheless,
there are terms in the equations that grow like $1/h^4$ at 
grid locations near the outer boundary, where $h$ is the
mesh spacing there. Therefore, for the equations to remain regular
near $i^0$ during evolution requires that the leading order behavior of 
the metric and scalar field variables always approach there asymptotic values
sufficiently fast to cancel this divergent behavior (this is essentially
the same problem one must deal with in an axisymmetric code near the
axis singularity). For the simple test results presented in Sec. \ref{sec_results}
the evolution near $i^0$ is well behaved, however we cannot guarantee that
this will be the case for all classes of asymptotically flat initial data.

A second issue is the propagation of outgoing waves toward $i^0$. The compactification
causes the wavelengths and speeds to decrease. Thus, for any fixed resolution
near $i^0$, such waves will eventually be poorly resolved on the grid\footnote{Keeping
the waves well resolved with AMR is not a practical solution in general, as the outgoing
wavetrain one expects from a binary inspiral, for example, is volume filling.}.
This could lead to a couple of undesirable effects. First, numerical dissipation
will significantly decrease the amplitude of the waves, making waveform extraction
in the outer regions of the domain impractical. Second, some portion of the
wave will get ``reflected'' back to the interior of the domain, which is not
physical and may adversely affect the accuracy of the interior solution.

\subsection{Apparent Horizon Finder and Excision}\label{sec_ahex}

We use the following {\em flow method} to search for single, simply-connected 
apparent horizons in the spacetime (this is the same algorithm used
in \cite{black_string}; see \cite{thornburg1} for a review of most current methods,
and \cite{thornburg2,schnetter} for some recent work on fast, elliptic-solver based 
apparent horizon finders). Consider the level set function
$F(r,\theta,\phi)$ defined by 
\be
F(r,\theta,\phi) = r - R(\theta,\phi),
\ee
where the spherical polar coordinates $(r,\theta,\phi)$ are defined
in terms of {\em uncompactified} coordinates, relative to some center
$(\bar{x}_0,\bar{y}_0,\bar{z}_0)$, via
\bea
\bar{x}&=&\bar{x}_0 + r \cos\phi\sin\theta \nonumber\\
\bar{y}&=&\bar{y}_0 + r \sin\phi\sin\theta \nonumber\\
\bar{z}&=&\bar{z}_0 + r \cos\theta
\eea
We want to find the function $R(\theta,\phi)$ such that the
hypersurface $F=0$ has zero outward null expansion $\Theta$
\be
\Theta = \ell_{\alpha;\beta} h^{\alpha\beta},
\ee
where $h^{\alpha\beta}$ is the spatial metric (\ref{adm})
and $\ell^\alpha$ is the outward pointing null vector normal to $F=\rm{const.}$
surfaces:
\be
\ell_\alpha = n_\alpha + \frac{h^\beta{}_\alpha \partial_\beta F}{\sqrt{h^{\delta\gamma} \partial_\delta F \partial_\gamma F}}.
\ee
The flow method involves specifying some initial guess for $R$, then
evolving the following equation until the magnitude of the norm of $\Theta$ evaluated 
along $F=0$ is as close to zero as desired:
\be\label{flow}
\frac{d R(\theta,\phi)}{d \lambda} = - \Theta(r,\theta,\phi)|_{r=R},
\ee
where $\Theta$ is evaluated along $F=0$.
This equation is parabolic in ``time'' $\lambda$, hence
$d\lambda$ must be of order $(\Delta x^i{})^2$ for stability.

During a typical evolution where a black hole forms via scalar
field collapse we initialize $R(\theta,\phi)=r_0$, where $r_0$ is a
constant close to though outside\footnote{The underlying assumption in
(\ref{flow}) is that $\Theta>0$ implies that the surface is outside
of the apparent horizon, which is {\em not} true everywhere at early times
during a gravitational collapse simulation.} of where we expect the apparent
horizon (AH) to first form, and periodically (every tens to hundreds of time steps) 
search for an AH using this initial guess until one is found. For subsequent
AH searches we use the previously found surface as an initial guess for $R(\theta,\phi)$.
If multiple black holes form we search for each AH independently.

Some form of excision is necessary for long term evolution of spacetimes
containing black holes. Excision means that one places interior boundaries
inside of all black holes such that all
physical singularities are removed from the computational domain. This
assumes that cosmic censorship holds, which further implies that a black
hole's event horizon will be {\em outside} any apparent horizon, and
hence one can use the apparent horizon as a guide where to excise.
For each black hole, we excise along an ellipsoid 
in {\em compactified} coordinate space, where the shape of the ellipsoid is chosen to match that 
of the apparent horizon as closely as possible along the ellipsoid's principal axis 
(which currently are required to lie along the coordinate axis). The size of the ellipsoid
is typically a bit smaller than that of the AH, to give some buffer zone between
the excision surface and the AH. Any point on the grid {\em inside} the ellipsoid
is defined to be excised, hence the excised region will necessarily
be a grid-based approximation to the smooth ellipsoidal shape (this is often
referred to as ``lego excision'' in the literature).

In general, boundary conditions need to be applied along the excision surface;
however, in a free evolution (such as described here) where all the
characteristics on the excision surface are directed inward, no boundary
conditions should be placed on the field variables. In the current
version of the code we {\em assume} that this is true, though we do 
not explicitly compute any of the characteristics. 
For a finite difference
scheme, such a ``no boundary'' boundary condition means that the evolutions
equations are applied at the excision surface, with centered difference
operators replaced, as appropriate, by forward or backward difference
operators so as not to reference grid values inside the excised region.
See Table \ref{ex_ops} for samples of the particular stencils
we use. Note that we define the excision surface to be constant in time,
and hence only spatial difference stencils need to be modified. During evolution,
if the excision surface moves such that previously excised points (interior points) become
``unexcised'', we initialize them via fourth order extrapolation from adjacent
exterior points at {\em all} time levels in the grid hierarchy.
We cannot {\em a-priori} prove that this excision method is stable,
rather, as discussed in Sec. \ref{sec_formalism}, we will require
convergence to a self-consistent solution of the field equations
as a proof-by-example that the code is stable and correct.

\begin{table}
\begin{tabular}[t]{| c || c |}
\hline
$f_{,x}$  & $(-3 f^n_i + 4 f^n_{i+1} - f^n_{i+2})/(2\Delta x)$ \\
          & \\
$f_{,xx}$ & $(2 f^n_i - 5 f^n_{i+1} + 4 f^n_{i+2} - f^n_{i+3})/(\Delta x)^2$ \\
\hline
\end{tabular}
\caption{A sample of the finite difference stencils used to convert
the differential equations to difference equations adjacent
to an excision surface. The column
on the right shows the second order accurate representation (with $y$ and
$z$ indices suppressed for clarity) of the corresponding derivative
operator to the left, evaluated at the point $(t^n,x_i,y_j,z_k)$.
The operators shown above are used when the 
point $x_{i-1}$ is inside the excision surface and the points $x_i,x_{i+1},...$ are 
outside of it.
}
\label{ex_ops}
\end{table}

\subsection{Matter Source}\label{sec_matter}
The present matter source modeled in the code is a massless scalar
field $\Phi$. The corresponding stress-energy tensor $T_{\mu\nu}$ is given by
\begin{equation}\label{set}
T_{\mu\nu} = 2 \Phi_{,\mu}\Phi_{,\nu} - g_{\mu\nu} \Phi_{,\gamma}\Phi^{,\gamma},
\end{equation}
and the evolution of $\Phi$ is governed by the wave equation
\begin{equation}\label{phi_eom}
\Box \Phi \equiv \Phi_{;\mu}{}^\mu = 0.
\end{equation}
Note that (\ref{set}) differs by a factor of $2$ from the convention of
Hawking and Ellis \cite{hawking_ellis}, which amounts to rescaling $\Phi$
by a factor of $\sqrt{2}$.

\subsection{Scalar field Initial Data}
At this stage for scalar field gravitational collapse we only consider time-symmetric 
initial data with a conformally
flat spatial metric.  Specifically, at $t=0$ the metric and its first time
derivatives take the following form:
\bea
\bar{g}_{tt}(t=0,x,y,z) &=& -1 \nonumber\\
\bar{g}_{ti}(t=0,x,y,z) &=& 0  \nonumber\\
\bar{g}_{ij}(t=0,x,y,z) &=& 0 \ \ \ , \ \ \ i\ne j \nonumber\\
\bar{g}_{ij}(t=0,x,y,z) &=& \Psi^4(x,y,z) \ \ \ , \ \ \ i=j \nonumber\\
\partial_t \bar{g}_{\alpha\beta}(t=0,x,y,z) &=& 0 \label{metric_ic}
\eea
The scalar field is thus the source of all non-trivial geometry 
at $t=0$, and we initialize it as a sum of Gaussian-like functions of the following
form
\bea
\Phi(t=0,x,y,z)  &=& \sum_i f^i(x,y,z), \nonumber\\
\partial_t \Phi(t=0,x,y,z) &=& 0, \label{phi_ic1}
\eea
with 
\bea
f^i(x,y,z) &=& A^i \exp\left(-[\rho^i(x,y,z)/\Delta^i]^2\right), \nonumber\\
\rho^i(x,y,z) &=& \{ [1-\epsilon^i_x{}^2][\bar{x}(x)-\bar{x}^i_0]^2 + \nonumber\\
       &\ & \ [1-\epsilon^i_y{}^2][\bar{y}(y)-\bar{y}^i_0]^2 + \nonumber\\
       &\ & \ [1-\epsilon^i_z{}^2][\bar{z}(z)-\bar{z}^i_0]^2 \}^{1/2}, \nonumber \\
\eea
where $A_i, \Delta^i, \epsilon^i_x, \epsilon^i_y, \epsilon^i_z, \bar{x}^i_0, \bar{y}^i_0$ 
and $\bar{z}^i_0$ are constants, and $\bar{x}(x),\bar{y}(y)$ and $\bar{z}(z)$ are given by (\ref{ct}).

In (\ref{metric_ic}), $\Psi(x,y,z)$ is solved for using the Hamiltonian constraint (\ref{hc})
equation, using an adaptive multigrid routine as discussed in \cite{fp_mgamr,fp_thesis}. 
Note however that some complications do arise when attempting to solve an elliptic
equation in compactified coordinates using multigrid; we will briefly discuss these issues
in Sec. \ref{sec_mg}.
The momentum 
constraints (\ref{mc}) are trivially satisfied with the above initial conditions. Once the constraints
have been solved, we initialize the source functions $\bar{H}_\alpha(t=0,x,y,z)$ using (\ref{barhdef})
and (\ref{hdef}).

With a three time level evolution scheme, the past time level at $t=-\Delta t$ needs to be
initialized as well. To obtain second order accurate convergence of the solution at late times, 
the past time level needs to be consistent with the initial data to within $\Delta t^2$. 
In the code we have implemented a couple of methods to achieve this; the first is to use
a Taylor expansion along with the equations of motion, the second is to evolve backward
in time to $t=-\Delta t$ with a smaller time step.
The first method works as follows\cite{matt_pc}.
For any one of the evolved grid functions $f^n_{ijk}$
the past time level $n=-1$ is initialized to second order accuracy 
using a Taylor expansion about $t=0$
\be
f^{-1}_{ijk} = f^0_{ijk} - f^{'0}_{ijk} \Delta t + f^{''0}_{ijk} \frac{\Delta t {}^2}{2},
\ee
where $f^{'0}_{ijk}$ is the first time derivative of $f(t,x,y,z)$ at $t=0$ (from the initial data),
and $f^{''0}_{ijk}$ is the second time derivative of $f(t,x,y,z)$ at $t=0$, evaluated by substituting
the initial data into the relevant equations of motion (\ref{efe_h},\ref{phi_eom}) and solving 
for $\partial_t\partial_t f$. For the second method, the past time level is only initialized to
first order using $f^0_{ijk}$ and $f^{'0}_{ijk}$, however with a smaller time step 
$\Delta t_s \approx \Delta t{}^2$. This initial data is then evolved backward in time until
$t=-\Delta t$, and the solution obtained there is used to initialize the past time level
for the actual evolution.

\subsubsection{Multigrid in a Compactified Coordinate System}\label{sec_mg}

Standard geometric multigrid (MG) methods that use point-wise relaxation as a smoother 
(as we do) are only efficient when the size of the coefficient functions
multiplying each of the principal parts of the elliptic operator are
of comparable size \cite{brandt}. This is the not the case near $i^0$ in
our compactified coordinate system. To illustrate, consider the 
form of the spatial Laplacian $\nabla^2$ using the coordinates of Sec. \ref{sec_coords}
in flatspace
\bea
\nabla^2 = &\ & \frac{4}{\pi^2} \Big{(}\cos^4(\pi x/2)\frac{\partial^2}{\partial x^2} + \nonumber
                                   \cos^4(\pi y/2)\frac{\partial^2}{\partial y^2} +  \\
         &\ & \ \ \ \ \ \          \cos^4(\pi z/2)\frac{\partial^2}{\partial z^2}\Big{)} + ... ,
%         &-& \frac{4}{\pi} \Big{(}\cos^3(\pi x/2)\sin(\pi x/2)\frac{\partial}{\partial x} + \nonumber \\
%         &\ & \ \ \ \ \ \           \cos^3(\pi y/2)\sin(\pi y/2)\frac{\partial}{\partial y} + \nonumber\\
%         &\ & \ \ \ \ \ \           \cos^3(\pi z/2)\sin(\pi z/2)\frac{\partial}{\partial z}\Big{)}
\eea
where the $...$ denote lower order terms. 
Notice that near any one of the outer boundaries $x^i=\pm 1$ the corresponding
coefficient of the second derivative term goes to zero. We have not solved
the issue of multigrid inefficiency in this part of the domain, however if
we use scalar field initial data of compact support in a sufficiently
small region about $(x,y,z)=(0,0,0)$, then we find that the fine-grid relaxation
performed within MG is adequate in obtaining a solution of sufficient accuracy
near $i^0$ (for $\Psi$ will then go like $1+O(1/r)$, and the initial guess of
$\Psi=1$ is close enough to the solution that relatively few relaxation sweeps
are needed).

A more serious problem is that relaxation using standard centered difference
approximations for first derivatives is unstable near $i^0$. A partial solution
to this problem is to use the following 4-way corner-averaged difference
operator at grid point $(i,j,k)$ (shown here for the $x$ derivative; the other first difference 
operators are similarly modified)
\bea
f_{,x} &=& \left(f_{i+1,j+1,k+1} - f_{i-1,j+1,k+1}\right)/(8\Delta x) + \nonumber\\
       &\ & \left(f_{i+1,j+1,k-1} - f_{i-1,j+1,k-1}\right)/(8\Delta x) + \nonumber\\
       &\ & \left(f_{i+1,j-1,k+1} - f_{i-1,j-1,k+1}\right)/(8\Delta x) + \nonumber\\
       &\ & \left(f_{i+1,j-1,k-1} - f_{i-1,j-1,k-1}\right)/(8\Delta x) \nonumber\\
       &\ & + O(\Delta x^2).
\eea
In the limit where the mesh spacing goes to zero in the vicinity of $x^i=\pm 1$, 
even this modification exhibits relaxation instabilities for initial data that
is not sufficiently compact about $r=0$. However, this is not a problem
for the kinds of physical systems we plan to use the code for, as all the interesting
dynamics will be confined to a small region abour $r=0$, and this will be the
part of the hierarchy with high resolution.

\subsection{Exact Schwarzschild Black Hole Initial Data}\label{plg_ic}

For some of the tests describe here we use analytic initial
data from a Schwarzschild black hole solution
in Painlev\'e-Gullstrand-like (PG) coordinates. The non-compactified
components of the metric are
\bea
ds^2 = &-&\left(1-\frac{2M}{\bar{r}}\right) dt^2 + d\bar{x}^2 + d\bar{y}^2 + d\bar{z}^2 \nonumber \\
       &+&\frac{2\sqrt{2M}}{\bar{r}^{3/2}}\left[\bar{x}d\bar{x} + \bar{y}d\bar{y} + \bar{z}d\bar{z}\right]dt,
\eea
where $\bar{r}\equiv\sqrt{\bar{x}^2+\bar{y}^2+\bar{z}^2}$ and $M$ is the mass of the black hole.
This (with appropriate spatial compactification) gives initial data
for the metric at $t=0$; and is also used to evaluate (\ref{hdef}) for the
initial values of the source functions.

\subsection{Efficient Simulation of Axisymmetric Spacetimes}\label{sec_axisym}
In this section a variant of the ``symmetry without symmetry'', or {\em Cartoon}
method \cite{cartoon} for efficient evolution of an axisymmetric spacetime with a 
Cartesian-based 3-dimensional code is described. The advantages to the
approach presented here are that {\em no} interpolation is ever performed,
the axisymmetric grid structure is a two dimensional
slice of the Cartesian grid, rather than a thin three dimensional slab,
and the method is not specific to finite difference based codes, so
can readily be applied to a spectral code, for instance.
Having the grid structure be two dimensional is helpful in that it allows easy
integration of the code with standard adaptive mesh refinement (AMR) packages.
The reason is as follows. In the original Cartoon algorithm, the third, thin
dimension is one finite-difference stencil-width thick.
However, most AMR algorithms can only refine a given
{\em volume} of a grid, which would increase the width of the slab-dimension
on finer levels, and thereby reduce the efficiency of the Cartoon method. Of coarse
the AMR algorithm could be modified to deal with such a situation, however
by using a two dimensional grid structure one avoids this problem altogether.
Note also that the purpose of the algorithm presented here is merely to provide
an efficient way to simulate axisymmetric spacetimes with a Cartesian code,
and {\em not} to address
any issues of axis stability in axisymmetric codes, which was one of the
original motivations behind Cartoon. Some recent work\cite{frauen} has suggested 
that the standard Cartoon algorithm may not be stable. Here we deal
with the axis by applying appropriate regularity conditions and numerical
dissipation, which has proven to be an effective method for dealing with
the axial singularity in axisymmetric codes\cite{garfinkle_axis,paper1}
(note also that in some cases stability in axisymmetric codes can be
obtained by constructing schemes with a conserved discrete energy, using
operators that satisfy summation by parts---see for example\cite{calabrese_neilsen}).

The idea behind our modified Cartoon algorithm is as follows. 
In a $d$-dimensional axisymmetric spacetime we have an azimuthal killing
vector $\xi^\mu$, hence the metric $g_{\mu\nu}$ and scalar field matter source $\Phi$ satisfy
\begin{eqnarray}
\mathcal{L}_\xi g_{\mu\nu}=0, \nonumber \\
\mathcal{L}_\xi \Phi=0. \label{kill_def}
\end{eqnarray}
What these equations imply is that all non-trivial structure of the metric and scalar
field are encoded within a $d-1$ dimensional sub-manifold $\mathcal{S}$ of the spacetime,
as long as $\xi^\mu$ is nowhere tangent to $\mathcal{S}$. Therefore, one
only needs to solve the field equations on $\mathcal{S}$, and 
(\ref{kill_def}) can be used to extend the solution throughout the spacetime.
In a numerical evolution, it makes most sense to have $\mathcal{S}$ 
coincide with a constant coordinate hypersurface, which we set to 
$\bar{z}=0$ for concreteness. We then choose coordinates such that $\xi^\mu$ has the
following explicit form (in uncompactified coordinates)
\begin{equation}\label{kill_vec}
\xi^\mu = \bar{y} \left(\frac{\partial}{\partial \bar{z}}\right)^\mu - 
          \bar{z} \left(\frac{\partial}{\partial \bar{y}}\right)^\mu
\end{equation}
which implies that $\xi^\mu$ is orthogonal to $\bar{z}=0$,
and the axis of symmetry runs along the $\bar{x}$ direction and is centered at
$\bar{y}=0$ \footnote{In other words, (\ref{kill_vec}) is merely the Cartesian
form of $(\partial/\partial\phi)^\mu$, where $\phi$ is a standard azimuthal
coordinate with the axis of symmetry coincident with the $\bar{x}$ axis.}.
To solve the field equations on $\bar{z}=0$ requires first
and second derivatives of metric variables both within the hypersurface
$\bar{z}=0$, and orthogonal to it in the $\bar{z}$ direction. To calculate
$\bar{z}$ derivatives the original Cartoon method effectively 
extends the solution using (\ref{kill_def}) to a sufficient number
of grid points above and below $\bar{z}=0$, so that the usual finite
difference stencils can be used to calculate $\bar{z}$ derivatives.
The approach taken here is to substitute the explicit form
of the Killing vector (\ref{kill_vec}) into the definition
(\ref{kill_def}), and use the resulting expression to evaluate
the $\bar{z}$ gradients directly. In other words, the same numerical method is used
to solve equations as outlined in Sec. \ref{sec_disc}, however instead
of calculating $\bar{z}$ derivatives using finite different approximation,
the  $\bar{z}$ derivatives are replace with appropriate combinations of
$\bar{x}$ and $\bar{y}$ gradients. In Appendix \ref{app_axi} we list the results
of this calculation for all relevant gradients of the metric in compactified
coordinates; here we illustrate the technique for the simpler case of the scalar field 
$\Phi$.

Evaluating (\ref{kill_def}) for $\Phi$ using (\ref{kill_vec}), we obtain
\begin{equation}\label{phi_z}
\frac{\partial \Phi}{\partial \bar{z}} = 
\frac{\bar{z}}{\bar{y}}\frac{\partial \Phi}{\partial \bar{y}} 
\end{equation}
Taking the $\bar{z}$ derivative of this equation, and replacing any $\bar{z}$ gradients
of $\Phi$ appearing on the right hand side with (\ref{phi_z}), gives
\begin{equation}\label{phi_zz}
\frac{\partial^2 \Phi}{\partial \bar{z}^2} = \frac{1}{\bar{y}}\frac{\partial \Phi}{\partial \bar{y}}
+ \frac{\bar{z}^2}{\bar{y}^2}\left(-\frac{1}{\bar{y}}\frac{\partial \Phi}{\partial \bar{y}}
+\frac{\partial^2 \Phi}{\partial \bar{y}^2}\right).
\end{equation}
Evaluating these equations at $\bar{z}=0$ gives
\begin{eqnarray}
\frac{\partial \Phi}{\partial \bar{z}}|_{\bar{z}=0} &=& 0, \nonumber \\
\frac{\partial^2 \Phi}{\partial \bar{z}^2}|_{\bar{z}=0} &=& \frac{1}{\bar{y}}\frac{\partial \Phi}{\partial \bar{y}} \label{phi_zzz0}
\end{eqnarray}
All other mixed second derivatives of $\Phi$ involving $\bar{z}$ are zero. 

One thing to note from equation (\ref{phi_zzz0}) is that the axis $\bar{y}=0$ is singular.
Therefore a regularity condition needs to be applied there, which can easily be 
seen from (\ref{phi_zzz0}) to be $\partial\Phi/\partial \bar{y}=0$ at $\bar{y}=0$. 

\section{Preliminary Results}\label{sec_results}
In this section we present results from several test simulations demonstrating certain
aspects of the code. Significantly more work needs to be done before the
code may be able to produce new physical results, however the current simulations
suggest that the generalized harmonic decomposition could be a viable alternative
to the ADM decomposition for many problems of interest.

In Sec. \ref{sec_cvf} we show a convergence test of scalar field evolution in 3D, 
Sec. \ref{sec_plg} evolves a Schwarzschild black hole in Painlev\'e-Gullstrand coordinates,
and Sec. \ref{sec_bh} demonstrates gravitational collapse of scalar field initial
data to a Schwarzschild black hole.

\subsection{Convergence Test}\label{sec_cvf}
For a 3D convergence test we used the following initial conditions for
$\Phi$ (\ref{phi_ic1}), which describes three prolate
spheroids slightly offset from one another so that there is no
spatial symmetry in the problem:
\bea
A^1&=&0.034, \ \ A^2=0.033, \ \ A^3=0.033 \nonumber\\
\Delta^1&=&0.1 \ \ \Delta^2=0.1 \ \ \Delta^3=0.1 \nonumber\\ 
(\bar{x}^1_0,\bar{y}^1_0,\bar{z}^1_0)&=&(0.025,0,0), \nonumber\\
(\bar{x}^2_0,\bar{y}^2_0,\bar{z}^2_0)&=&(0,-0.025,-0.025), \nonumber\\
(\bar{x}^3_0,\bar{y}^3_0,\bar{z}^3_0)&=&(-0.025.025,0.025) \nonumber\\
\epsilon^1_x&=&0.1, \ \ \epsilon^2_y=0.1, \ \ \epsilon^3_z=0.1 
\eea
and all other initial data parameters for $\Phi$ are zero. 
With these parameters the ADM mass of the spacetime is roughly
$0.005$, so the initial distribution of energy is concentrated
in a radius about 10 times larger than its effective Schwarzschild radius.
For the $H_t$ coordinate condition we used a slightly modified version of (\ref{box_hdotn}),
where we eliminated the coupling (through the normal $n^\mu$) to $H_i$
and added an arbitrary power $n$ of $\alpha$ in the denominator:
\be
\Box H_t = - \kappa_t(t) \frac{\alpha-\alpha_0}{\alpha^n} + 
            \xi_t(t) H_t{}_{,\mu} n^\mu \label{box_hdotn_b},
\ee
where $\kappa_t(t)=\kappa_0 q(t)$, $\xi_t(t)=\xi_0 q(t)$, and
$q(t)$ is given by
\bea\label{qt}
q(t) &=& \left(\frac{t}{t_1}\right)^3 
         \left[6 \left(\frac{t}{t_1}\right)^2 - 15 \frac{t}{t_1} + 10 \right], \ \ \ 0\le t\le t_1 \nonumber \\
     &=& 0, \ \ \ {\rm elsewhere.} 
\eea
$q(t)$ provides a smooth (twice differentiable) transition from $0$ at $t=0$
to $1$ at $t=t_1$, and makes the evolution of the source functions
consistent with the choice of time-symmetric initial data.
We evolve $\bar{H}_i$ to zero using a version of 
(\ref{harm_dgauge}) with $H_i$ replaced by $\bar{H}_i$, and
$\kappa_i(t)=\kappa_0 q(t)$.
For this particular simulation we had $\kappa_0=50$, $\xi_0=10$, $n=3$, $t_1=1/10$
and $\alpha_0=1$.

\begin{figure}
\begin{center}
\includegraphics[width=3.75cm,clip=true]{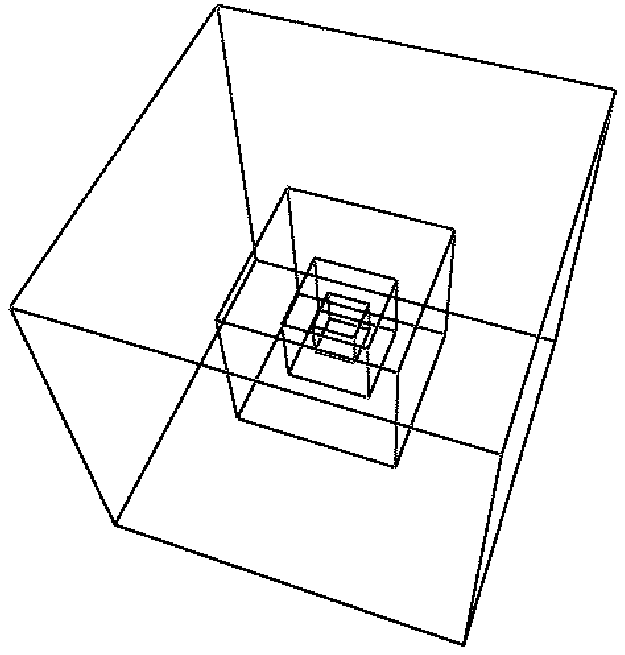}
\includegraphics[width=3.75cm,clip=true]{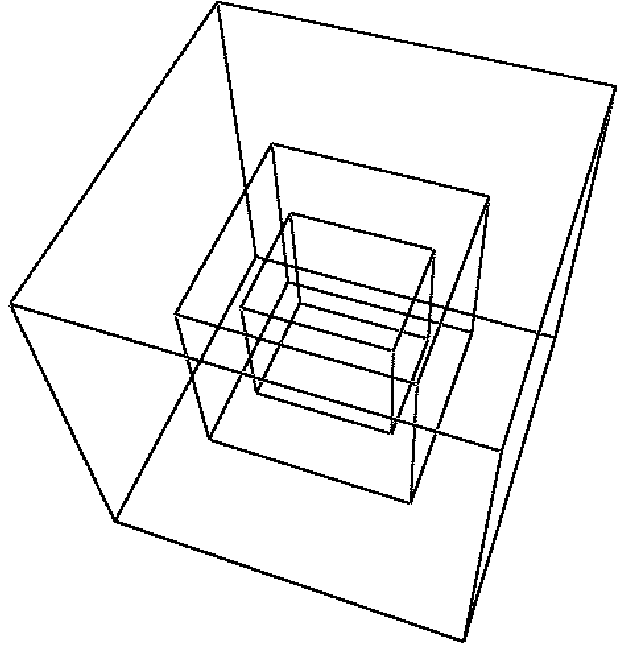}
\end{center}
\caption {A depiction of the adaptive mesh structure for the convergence
test simulation described in Sec. \ref{sec_cvf}. The image to the
left corresponds to the mesh structure at $t=0$, while that to the
right at $t=0.5$. The largest box in each figure, whose faces are
at $i^0$, actually represent two levels of (2:1) refinement. The increase
in size of the finer levels and loss of the finest level of refinement
by $t=0.5$ is due to the outward propagation of the initial distributions of
energy.
}
\label{sample_amr_mesh}
\end{figure}

A convergence test involves running a given simulation at several
different resolutions, and comparing the results to ensure that
the solution of the finite difference equations is converging
to a solution of the partial differential equations. 
We ran three simulations of differing resolution. The coarsest
resolution run had a base grid size of $17^3$, and we specified a 
value for the maximum desired truncation error so that up to $5$
additional levels of $2:1$ refinement were used, giving an {\em effective} 
finest resolution of $513^3$---see Fig. \ref{sample_amr_mesh} below for
a depiction of the mesh structure at two times during the simulation. 
We used a Courant factor of $0.25$ at
each level in the hierarchy (i.e. $\Delta t = 0.25 \Delta x^i$).
For the medium and finest resolution simulations we used the same 
grid hierarchy produced
by the coarsest resolution simulation (which was produced using standard
truncation error estimate methods), though
doubled and quadrupled the resolution of all the grids respectively, 
keeping the same Courant factor. 
To keep the computational cost of the highest resolution run manageable, we
only ran the simulation until $t=0.5$; however this corresponds to roughly 
five light-crossing
times of the central region of the grid where the scalar field is concentrated,
and so a reasonable amount of dynamics does occur. Also, this run time is sufficiently
long that possible adverse effects from the AMR algorithm, such as from regridding or
high-frequency ``noise'' from parent-child refinement boundaries, can be captured
by the convergence test.

\begin{figure}
\begin{center}
\includegraphics[width=8.5cm,clip=true]{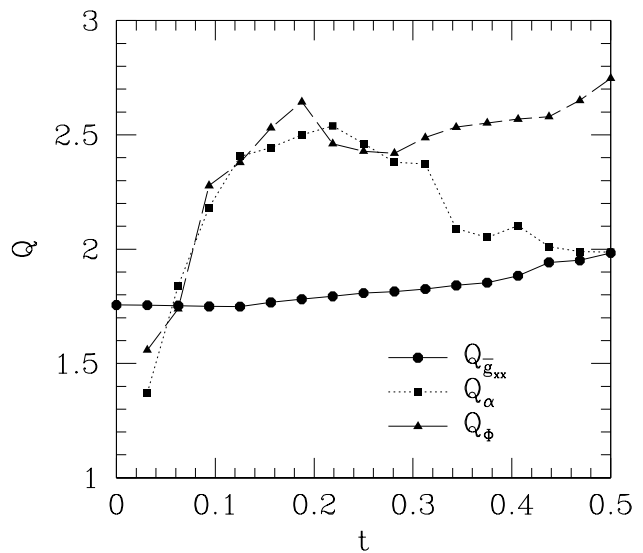}
\end{center}
\caption{
Convergence factors (\ref{Qdef}) for representative
grid functions from the simulation described in Sec. \ref{sec_cvf}.
The points denote the times when $Q$ was calculated, and correspond
to times when the entire grid hierarchy was in sync. Note that we only
show $Q_{\bar{g}_{xx}}$ at $t=0$, as all the other functions are exactly
known then, and hence $Q$ is ill-defined. This plot shows that the solution
is close to second order convergent, with some caveats discussed in the text.
\label{cvf_rest}}
\end{figure}

Label some grid function $f$ from the finest resolution 
simulation $f_h$, from the medium one $f_{2h}$ and from the coarsest one $f_{4h}$.
Then the convergence factor $Q_f$ we calculate is
\be\label{Qdef}
Q_f = \frac{1}{\ln 2} \left(\ln\|f_{4h}-f_{2h}\|-\ln\|f_{2h}-f_{h}\|\right),
\ee
where before the subtraction we interpolate the grid functions to a common
uniform grid, and then compute the $\ell_2$ norm of the differences. For an $n^{th}$ order accurate
scheme one would expect $Q_f$ to approach a value of $n$ in the limit as
the mesh spacing goes to zero. See Fig. \ref{cvf_rest} for the convergence
factors from the above simulations for several representative functions. 
The plot shows that we do see convergence close to second order. At
early times, the convergence factor is slightly worse than second order;
we surmise that the reason for this is a small amount of 
unphysical, high-frequency solution components (``noise'') present
at parent-child mesh refinement 
boundaries at the initial time. This noise seems to come from the
multigrid algorithm we use to solve for the initial data,
where linear interpolation is used to prolong from the coarse to fine
meshes.
Linear interpolation introduces
high-frequency components in the fine grid solution, which is
smoothed by relaxation, however relaxation is only applied at interior
points. Presumably some form of explicit dissipation at parent child
boundaries, or higher order interpolation could cure this problem, 
though we find that the dissipation
we use during subsequent evolution is also quite effective at reducing the
magnitude of this noise. At late times, several grid functions seem to show
anomalously high convergence factors. This seems to be due to the fact
that the simulations (in particular the coarsest resolution one)
are not yet that close to the convergent regime. To test this would require
a higher resolution simulation, which would be impractical because of our computer
resource limitations\footnote{Alternatively, we
can choose initial data that is better resolved on the coarsest grid; however,
the kind of resolution we have here is more representative of the resolution
we will be able to achieve in the near future with the computer power
we have access to, and so we think this is a fair test of the code.}. However, by
looking at an independent residual of the Einstein equations, as described 
next, we can already see the {\em trend} towards second order convergence using
only three simulations.

\begin{figure}
\begin{center}
\includegraphics[width=8.5cm,clip=true]{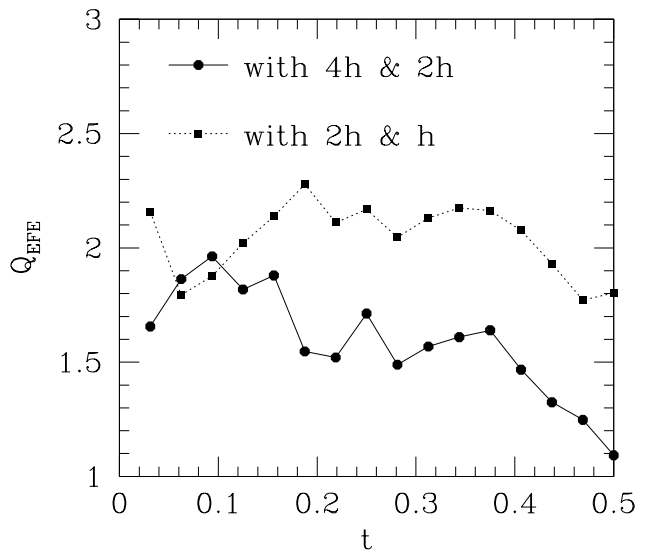}
\end{center}
\caption {
Convergence factor of the independent residual (\ref{ires},\ref{Qrdef}) 
of the Einstein equations from the simulation described in Sec. \ref{sec_cvf}.
The points denote the times when $Q_\mathcal{R}$ was calculated, and corresponds
to times when the entire grid hierarchy is in sync {\em after} an
evolution time step (hence there are no points at $t=0$).
This plot shows we are tending
toward a solution that is second order convergent.
\label{cvf_efe}}
\end{figure}

To check that we are solving the Einstein equations we compute an 
independent residual $\mathcal{R}_{\alpha\beta}$ of (\ref{efe}) 
\be\label{ires}
\mathcal{R}_{\alpha\beta} = R_{\alpha\beta}-4\pi\left(2\pi T_{\alpha\beta}-g_{\alpha\beta} T\right).
\ee
After discretizing the ten
residuals $\mathcal{R}_{\alpha\beta}$ using the finite difference
stencils described in the preceding section, we
compute the residual grid function $\mathcal{R}$ at each 
grid point as the infinity norm over the ten residuals.
Note that we compute (\ref{ires}) {\em without} reference to the source functions,
using only the {\em compactified} metric elements and scalar field.
Since we know that $\mathcal{R}$ should converge to zero in the limit,
it is sufficient to compute its convergence factor using two resolutions, 
for example
\be\label{Qrdef}
Q_\mathcal{R} = \frac{1}{\ln 2} \left(\ln\|\mathcal{R}_{2h}\|-\ln\|\mathcal{R}_{h}\|\right).
\ee
Fig. \ref{cvf_efe} shows $Q_\mathcal{R}$ computed using
both $[\mathcal{R}_{2h},\mathcal{R}_{h}]$ and $[\mathcal{R}_{4h},\mathcal{R}_{2h}]$.
This plot shows that we are tending towards second order convergence as the
resolution is increased.

\subsection{Schwarzschild Black Hole Evolution}\label{sec_plg}

In this section we briefly show how well the current code can evolve
a Schwarzschild black hole in Painlev\'e-Gullstrand coordinates.
The analytic solution is used for initial conditions as described
in Sec. \ref{plg_ic}, with $M=0.05$, and using (\ref{harm_gauge}) to keep
the source functions frozen-in during evolution. A (``lego'') spherical
excision region of radius $1.2M$ was used. 
We ran three axisymmetric simulations,
each with identical grid hierarchy, though successively higher resolution
as described in the previous section. The lowest resolution simulation
had a base grid of $33\mbox{x}17$ (spanning $-1..1$ in $x$ and $0..1$ in $y$),
using 6 additional levels of 2:1 refinement, and a Courant factor of $0.125$
(the mesh structure is very similar to that depicted in Fig.\ref{sample_amr_mesh},
however in this example the refinement is constant in time). 
To compare, we also ran the two lowest resolution simulations of the 
equivalent problem in full
3D; lack of computational resources prevented us from running
the highest resolution simulation in 3D, and, for the same reason,
we were not able to run the
medium resolution simulation as long as the axisymmetric 
one\footnote{Specifically, the medium resolution ($2h$) simulation
in 3D took 160 hours of runtime on 128 nodes of the Westgrid Xeon cluster
to reach $t=55M$, using about 120MB of memory on each node. By comparison,
the highest resolution ($h$) axisymmetric simulation took approximately
240 hours on 24 nodes of UBC's vn4 Xeon cluster to reach $t=220M$.
In 3D (2D), doubling the resolution typically requires 16(8) times
the runtime, and 8(4) times the memory to evolve to a given
physical time in the simulation.}.

\begin{figure}
\begin{center}
\includegraphics[width=8.5cm,clip=true]{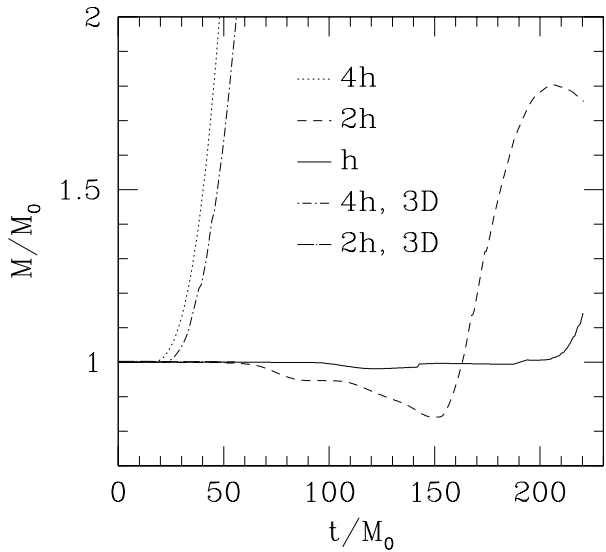}
\end{center}
\caption{
Normalized mass (\ref{ah_mass}) for the evolution of a vacuum 
$M_0=0.05$ Schwarzschild black hole in Painlev\'e-Gullstrand coordinates.
The curve labeled $4h$ corresponds to the lowest resolution axisymmetrix
simulation, while the $2h$ ($h$) curves are from axisymmetric 
simulations with twice (four times) the resolution. The curves $4h, 3D$ and
$2h, 3D$ are from runs with identical resolution to the $4h$ and $2h$ axisymmetric
simulations respectively, though the simulations were in full 3D. Note that
the $2h, 3D$ simulation curve only extends till roughly $t/M_0=55$ (as we ran out of 
computer time then), and is effectively hidden behind the
other curves as $M/M_0\approx 1$ up till then.
\label{plg_m}}
\end{figure}

As a measure of the accuracy of the simulation, we calculate the mass $M$
of the black hole from the area $A$ of the apparent horizon: 
\be\label{ah_mass}
M = \sqrt{\frac{A}{16\pi}}.
\ee
The mass for the five simulations is shown in Fig. \ref{plg_m}. 
Note that we have not calculated any errors associated with the
numerical integration of apparent horizon area; we used the
same resolution sphere ($33$ points in $\theta$, ranging from $0$ to $\pi$)
in all cases, hence the error in the area calculation will be roughly the same
for each run.
There are a couple of significant things to note from this figure.
First, even though we were not
able to fully compare the axisymmetric results with a 3D code, what
the partial comparison {\em suggests} is that explicitly enforcing axisymmetry
in this case does {\em not} have a significant effect on the accuracy or runtime
of the simulation.
Second, even though the length of time that we can simulate a black hole
to within a given accuracy with this code is not too long compared 
to the state of the art these days,
the trend in increased run-time with resolution is promising. In particular,
there is not much evidence of exponential growth of error early on (though once
the error has grown to a certain magnitude, the code crashes quickly); rather, these
plots {\em suggest} that the leading order truncation error term has somewhere between
linear and quadratic dependence on time. To see this, let us compare
the putative runtime of a simulation where the leading order error grows exponentially
with time, versus polynomial growth of the error. For the exponential 
case, assume the norm of the error $E(t)$ as a function of time takes the following form
\be
E(t)=C h^2 e^{\lambda t},
\ee
where $C$ is some constant, $\lambda$ is the continuum growth factor, and
$h$ is the mesh spacing. In other words, this situation describes an
exponential ``constraint violating mode'' driven by truncation error terms.
Let us solve for the evolution time $t_h$, to reach a specified error $E(t)=E_0$
with mesh spacing $h$:
\be\label{tl}
t_h = \frac{\ln E_0 - \ln C -2\ln h}{\lambda}
\ee
Now consider the following quantity $\zeta$ computed using three simulations
with differing resolutions:
\be
\zeta \equiv \frac{t_h - t_{2h}}{t_{2h} - t_{4h}}
\ee
Evaluating $\zeta$ for the case of exponential growth using (\ref{tl}) gives
\be
\zeta_\lambda = 1.
\ee
Repeating the calculation for the case of polynomial error growth of the form
\be
E(t)=C h^2 t^p
\ee
gives
\be
\zeta_p = \frac{2^{2/p}-1}{1-2^{-2/p}}.
\ee
For linear error growth, $\zeta_{p=1}=4$, for quadratic growth $\zeta_{p=2}=2$,
and $\zeta_p\rightarrow 1$ in the limit as $p\rightarrow \infty$.
If we evaluate $\zeta$ by defining the error to be that in $M/M_0$ from
Fig.\ref{plg_m}, using a value of $3\%$ for $E_0$ we compute $\zeta\approx2.7$,
suggesting polynomial rather than exponential growth. However, this
number changes as $E_0$ changes (for example setting $E_0$ to $10\%$ 
suggests faster than exponential growth), so we cannot conclusively
rule out exponential growth. Regardless, from the practical point of
view of using the current code to investigate black hole physics in
3D, we need prohibitively high resolutions to get to a useful
runtime range of several hundred $M$, so significantly more work
needs to be done to improve the code for black hole simulations.

\subsection{Black Hole Formation}\label{sec_bh}
\begin{figure}
\begin{center}
\includegraphics[width=8.5cm,clip=true]{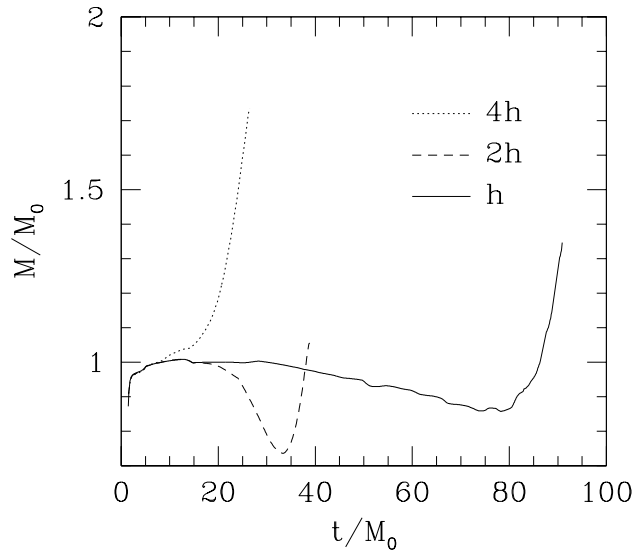}
\end{center}
\caption{
Normalized mass (\ref{ah_mass}) from the axisymmetric 
evolution of a black hole
formed via the gravitational collapse of a scalar field. 
The value of $M_0$ used was the largest, convergent value
of $M$ from the three simulations, which is a reasonable estimate
of the final mass of the black hole. For comparison, the
grid structure used for the simulations was identical to the
corresponding axisymmetric simulations shown in Fig.\ref{plg_m}
\label{axi_bh_m}}
\end{figure}

The final test presented here is gravitational collapse of scalar field
initial data to a black hole, in axisymmetry. To compare with the vacuum black 
hole simulation of the previous section, we used an identical grid structure,
and chose initial data so that a black hole of roughly the same
mass ($0.05$) forms. 
Specifically, we used a spherically symmetric Gaussian
pulse (\ref{phi_ic1}) with 
\be
A^1 = 0.35, \ \ \ \Delta^1 = 0.055,
\ee
with the rest of the initial data parameters for $\Phi$ set to zero.
We used the same gauge conditions for $H_\mu$ as described in 
Sec. \ref{sec_cvf} for the convergence test, except here the
corresponding parameters were $\kappa_0=40$, $\xi_0=30$, $n=5$, $t_1=1/80$
and $\alpha_0=1$. Note that the results are not very sensitive to this
particular choice of gauge parameters; the rule of thumb is
that $\kappa_0$ and $\xi_0$ of order $1/\Delta^1$, $t_1$ 
of order $\Delta^1$ and $n$ of order unity works reasonably well.
Fig. \ref{axi_bh_m} shows the corresponding plot of apparent horizon
mass versus time. The black hole forms after about $2M$ of
evolution, after which some accretion of scalar field occurs, causing the
mass to grow by a bit early on. 
Note also that once we detect an apparent horizon, we
excise a spherical region $60\%$ the size of the horizon, so approximately
at $1.2M$, again for comparison with the previous evolution.
At a given resolution this 
simulation (as judged by the mass estimate) has less accuracy
compared to the corresponding vacuum simulation, however the trend
of increased accuracy with increased resolution is roughly the same.

\section{Conclusion}\label{sec_conclusion}

We have described a new computational scheme for numerically solving the
Einstein equations based upon generalized harmonic coordinates. 
This extends earlier work of Garfinkle \cite{garfinkle}, and in some respects is
similar to the direction pursued by Szilagyi and Winicour \cite{szilagyi_winicour}.
Some of the topics covered included suggestions for imposing dynamical
gauge conditions, a new technique of implementing the Cartoon method \cite{cartoon}
for simulating axisymmetric spacetimes with a Cartesian code, 
a direct discretization scheme for second-order-in-space-and-time
partial differential equations, and the use
of a spatially compactified coordinate system. 
One attractive feature of harmonic evolution is that the principal
part of the Einstein equations reduce to wave equations for each metric element.
This, together with the use of a second order discretization scheme,
keeps the number of variables and constraint equations to a minimum,
and the hope is that this will make it easier to achieve stable evolution.
The use of a spatially
compactified domain allows one to impose correct asymptotic boundary
conditions for the simulation, and thus we automatically have constraint
preserving boundary conditions.
The advantage of our Cartoon method over the original is that no interpolation
is needed, and the simulation is performed on a 2D slice of the spacetime, 
thus simplifying the process of incorporating the code into an adaptive
mesh refinement framework. Furthermore, the technique is not particular
to finite difference codes, and can be used with spectral methods, for instance.

Preliminary test simulations of black hole spacetimes suggest that
this scheme holds promise for being applicable to many problems
of interest, including the binary black hole problem, black hole-matter interactions,
and critical gravitational collapse.
However, a lot of research still
needs to be done, at both the analytical and numerical levels,
before this scheme may produce new physical results. In particular,
it would be useful to analyze the mathematical well-posedness of the
fully discrete system, including a variety of possible gauge evolution
equations. The majority of techniques for analyzing 
hyperbolic systems require reduction to first order form
(recently similar techniques have been developed for second order
in space, first order in time systems \cite{nagy_et_al,gundlach_garcia_1,
gundlach_garcia_2}; also, in \cite{beyer_sarbach} the BSSN system is
analyzed by converting to first order form, however the constraints introduced
by this reduction are shown to obey a closed evolution system that is independent
of the other constraints, implying that the original second order system
is well-posed). 
At the numerical level,
a broader class of initial conditions needs to be explored, such as
black hole/matter interactions and black hole collisions. This is of course
one of the primary long term goals of the code, however early tests indicate
that a significant number of adjustments and improvements (to dissipation
and extrapolation operators, for example) may be needed, in addition
to more sophisticated gauge conditions than discussed here, before
such scenarios could be simulated with sufficient accuracy and length
of time for new results to be obtained.

\section*{Acknowledgments}
I would like to thank Matthew Choptuik for many stimulating discussions about
the work presented here.
I gratefully acknowledge research support from
NSF PHY-0099568, NSF PHY-0244906 and Caltech's Richard Chase Tolman Fund.
Simulations were performed on UBC's {\bf vn}
cluster, (supported by CFI and BCKDF), and the {\bf Westgrid} cluster
(supported by CFI, ASRI and BCKDF).

\appendix

\section{Evolution of Axisymmetric Spacetimes}\label{app_axi}
As described in Sec. \ref{sec_axisym}, we can efficiently simulate
axisymmetric spacetimes along a single $z=0$ slice of the spacetime
by replacing all $z$ gradients in the field and matter evolution
equations with appropriate $x$ and $y$ gradients, as dictated
by (\ref{kill_def}).
Here we list the equations for gradients of the {\em regular}
components of the metric $g_{\mu\nu}$ and scalar field $\Phi$ with respect to 
the {\em compactified} coordinates (Sec. \ref{sec_coords}), and
give the corresponding on-axis regularity conditions.

The first $z$ derivatives are:
\bea
\partial_z\bar{g}_{tt}|_{z=0} &=& 0 \nonumber \\
\partial_z\bar{g}_{t\bar{x}}|_{z=0} &=& 0 \nonumber \\
\partial_z\bar{g}_{t\bar{y}}|_{z=0} &=& - \bar{g}_{t\bar{z}}(\pi/2\bar{y}) \nonumber \\
\partial_z\bar{g}_{t\bar{z}}|_{z=0} &=& \bar{g}_{t\bar{y}}(\pi/2\bar{y}) \nonumber \\
\partial_z\bar{g}_{\bar{x}\bar{x}}|_{z=0} &=& 0 \nonumber \\
\partial_z\bar{g}_{\bar{x}\bar{y}}|_{z=0} &=& - \bar{g}_{\bar{x}\bar{z}}(\pi/2\bar{y}) \nonumber \\
\partial_z\bar{g}_{\bar{x}\bar{z}}|_{z=0} &=& \bar{g}_{\bar{x}\bar{y}}(\pi/2\bar{y}) \nonumber \\
\partial_z\bar{g}_{\bar{y}\bar{y}}|_{z=0} &=& - \bar{g}_{\bar{y}\bar{z}}(\pi/\bar{y}) \nonumber \\
\partial_z\bar{g}_{\bar{y}\bar{z}}|_{z=0} &=& (\bar{g}_{\bar{y}\bar{y}}-\bar{g}_{\bar{z}\bar{z}})(\pi/2\bar{y}) \nonumber \\
\partial_z\bar{g}_{\bar{z}\bar{z}}|_{z=0} &=& \bar{g}_{\bar{y}\bar{z}}(\pi/\bar{y}) \nonumber \\
\partial_z\Phi|_{z=0} &=& 0 \label{axid}
\eea
Mixed $z-t$, $z-x$ and $z-y$ second derivatives are calculated
by taking the appropriate derivative of (\ref{axid}). Second
derivatives with respect to $z$ are computed as follows:
\bea
\partial_z\partial_z\bar{g}_{\alpha\beta}|_{z=0} &=& 
\frac{\pi}{2}\left(\frac{\partial_y\bar{g}_{\alpha\beta}}{\bar{y}(1+\bar{y}^2)}
                     +\frac{\pi C_{\alpha\beta}}{2\bar{y}^2} \right)\nonumber \\
\partial_z\partial_z\Phi|_{z=0} &=& \frac{\pi\partial_y\Phi}{2\bar{y}(1+\bar{y}^2)} \label{axidd1}
\eea
where the coefficients $C_{\alpha\beta}$ are
\bea
C_{tt} &=& 0 \nonumber \\
C_{t\bar{x}} &=& 0 \nonumber \\
C_{t\bar{y}} &=& - \bar{g}_{t\bar{y}} \nonumber \\
C_{t\bar{z}} &=& - \bar{g}_{t\bar{z}} \nonumber \\
C_{\bar{x}\bar{x}} &=& 0 \nonumber \\
C_{\bar{x}\bar{y}} &=& - \bar{g}_{\bar{x}\bar{y}} \nonumber \\
C_{\bar{x}\bar{z}} &=& - \bar{g}_{\bar{x}\bar{z}} \nonumber \\
C_{\bar{y}\bar{y}} &=& 2 (\bar{g}_{\bar{z}\bar{z}}-\bar{g}_{\bar{y}\bar{y}}) \nonumber \\
C_{\bar{y}\bar{z}} &=& - 4 \bar{g}_{\bar{y}\bar{z}} \nonumber \\
C_{\bar{z}\bar{z}} &=& 2 (\bar{g}_{\bar{y}\bar{y}}-\bar{g}_{\bar{z}\bar{z}})
\eea
The on axis regularity conditions are
\bea
\partial_y\bar{g}_{tt}|_{y=0} &=& 0 \nonumber \\
\partial_y\bar{g}_{t\bar{x}}|_{y=0} &=& 0 \nonumber \\
\bar{g}_{t\bar{y}}|_{y=0} &=& 0 \nonumber \\
\bar{g}_{t\bar{z}}|_{y=0} &=& 0 \nonumber \\
\partial_y\bar{g}_{\bar{x}\bar{x}}|_{y=0} &=& 0 \nonumber \\
\bar{g}_{\bar{x}\bar{y}}|_{y=0} &=& 0 \nonumber \\
\bar{g}_{\bar{x}\bar{z}}|_{y=0} &=& 0 \nonumber \\
\partial_y\bar{g}_{\bar{y}\bar{y}}|_{y=0} &=& 0 \nonumber \\
\partial_y\bar{g}_{\bar{y}\bar{z}}|_{y=0} &=& 0 \nonumber \\
\partial_y\bar{g}_{\bar{z}\bar{z}}|_{y=0} &=& 0 \nonumber \\
\partial_y\Phi|_{y=0} &=& 0 \nonumber \\
\bar{g}_{\bar{y}\bar{y}}|_{y=0} &=& \bar{g}_{\bar{z}\bar{z}}|_{y=0}
\eea

To compute the $z$ gradients and regularity conditions for the
source functions in the code, we simply substitute the results from
the calculation for the metric into the definition of the source functions
(\ref{hdef},\ref{barhdef}). 

\section{Stability Analysis of a Second Order in Space and Time Evolution Scheme}\label{app_stab}

Here we give a von Neumann-like stability analysis of the one dimensional flat space wave 
equation using the discretization scheme described in Sec. \ref{sec_code}.
Second order in time schemes for the wave equation are not very common in the
literature, so the example given here is to demonstrate that the method is
inherently stable, ignoring the complications of boundaries, excision,
non-constant coefficients and non-linear lower order terms of the full problem
(the analysis of which is beyond the scope of this paper). Even
though dissipation is not needed in this example, we add it as applied in the
code to demonstrate how it works.

The model wave equation for $\Phi(x,t)$ is
\begin{equation}
\Phi_{,tt} - \Phi_{,xx}=0.
\end{equation}
Discretization of this equation using the stencils in Tab. \ref{ops} gives
\begin{equation}
\Phi^{n+1}_j -2 \Phi^{n}_j + \Phi^{n-1}_j - 
\lambda^2\left(\Phi^{n}_{j+1} -2 \Phi^{n}_j + \Phi^{n}_{j-1}\right)=0,
\end{equation}
where $\Phi^n_j\equiv\Phi(x=j\Delta x,t=n\Delta t)$ and $\lambda\equiv\Delta t/\Delta x$ is
the Courant factor. This immediately gives an explicit update scheme
for the unknown $\Phi^{n+1}$ given information at two past time levels, 
$\Phi^{n}, \Phi^{n-1}$:
\begin{equation}\label{phi1ddisc}
\Phi^{n+1}_j = 2 \Phi^{n}_j - \Phi^{n-1}_j + 
\lambda^2\left(\Phi^{n}_{j+1} -2 \Phi^{n}_j + \Phi^{n}_{j-1}\right)
\end{equation}
(note that the iterative relaxation method described
in Sec. \ref{sec_code} gives exactly the same update scheme in this case).
It is mathematically simpler to analyze this equation using an equivalent
two time level scheme by introducing the variable
\begin{equation}
\Psi^n_j \equiv \Phi^{n-1}_j,
\end{equation}
after which (\ref{phi1ddisc}) becomes
\begin{eqnarray}
\Phi^{n+1}_j &=& 2 \Phi^{n}_j - \Psi^{n}_j + 
\lambda^2\left(\Phi^{n}_{j+1} -2 \Phi^{n}_j + \Phi^{n}_{j-1}\right) \nonumber\\
\Psi^{n+1}_j &=& \Phi^{n}_j. \label{phi1ddisc2}
\end{eqnarray}
As (\ref{phi1ddisc2}) is linear with constant coefficients, we can completely
characterize its stability properties by analyzing the evolution of 
individual Fourier modes of the form $c(t) e^{i k x}$. To this end, let
\begin{eqnarray}
\Phi(x,t) \equiv a(t) e^{i k x} \\
\Psi(x,t) \equiv b(t) e^{i k x}.
\end{eqnarray}
Substituting this into (\ref{phi1ddisc2}) gives
\begin{eqnarray}
a^{n+1} &=& 2 a^n - b^{n} - 4 \lambda^2 \xi^2 a^n\nonumber\\
b^{n+1} &=& a^n, \label{phi1ddisc3}
\end{eqnarray}
where
\begin{equation}
\xi\equiv\sin(k\Delta x/2),
\end{equation}
and we have used the identity 
$-4\sin^2(k\Delta x/2) = e^{- i k \Delta x} -2 + e^{ i k \Delta x}$.
Note that the smallest wavelength that can be represented on a
numerical grid is $2\Delta x$ (the Nyquist limit), which corresponds to
a largest possible wave number $k=\pi/\Delta x$, hence $\xi$ ranges from $0$ to $1$.

As described in Sec. \ref{sec_diss}, we apply numerical dissipation
to all past time level variables, prior to the update step, by first
calculating the high-frequency component of the function using (\ref{x_hf}),
and then subtracting it from the function via (\ref{x_hf_sub}). 
For a grid function $f^n_j=c^n e^{i k x_j}$, the high-frequency component $\eta^n_j$ is
defined as
\begin{eqnarray}
\eta^n_j &=& \frac{1}{16}\left(f^n_{j-2} - 4f^n_{j-1} + 6 f^n_j  - 4f^n_{j+1} + f^n_{j+2}\right)\nonumber\\
         &=& f^n_j \xi^4,
\end{eqnarray}
and filtering amounts to modifying $f^n_j$ as follows
\begin{eqnarray}
f^n_j &\rightarrow& f^n_j - \epsilon \eta^n_j \nonumber\\
      &=& f^n_j (1 - \epsilon \xi^4)\nonumber\\
      &=& \bar{\epsilon} f^n_j,
\end{eqnarray}
where $\bar{\epsilon}\equiv 1 - \epsilon \xi^4$. As $\xi\in[0..1]$ and $\epsilon\in[0..1]$, 
$\bar{\epsilon}\in[0..1]$. With this form of 
dissipation (which is linear, and hence fits 
into the Fourier analysis of the evolution scheme) applied to both 
$\Phi^n_j$ and $\Psi^n_j$, (\ref{phi1ddisc3}) becomes
\begin{eqnarray}
a^{n+1} &=& \bar{\epsilon} \left[2 a^n\left(1 - 2 \lambda^2 \xi^2\right) - b^{n}\right] \nonumber\\
b^{n+1} &=& \bar{\epsilon} a^n \label{phi1ddisc4}.
\end{eqnarray}
In matrix form, the update step can be written as
\begin{equation}
\left[\begin{array}{c}a \\ b \end{array}\right]^{n+1} = {\bold A}
\left[\begin{array}{c}a \\ b \end{array}\right]^{n},
\end{equation}
where
\begin{equation}
{\bold A} = \bar{\epsilon} \left[\begin{array}{cc} 2 \left(1 - 2 \lambda^2 \xi^2\right) & -1 \\ 1 & 0 \end{array}\right].
\end{equation}
The numerical evolution will be stable if the eigenvalues $\Lambda_\pm$ of ${\bold A}$ all
lie on or within the unit circle in the complex plain. A straight-forward calculation gives
\begin{equation}
\Lambda_\pm = \bar{\epsilon} \left[ 1 - 2\xi^2\lambda^2 \pm i 2 \xi\lambda \sqrt{1 - \xi^2\lambda^2}\right]
\label{eigs}
\end{equation}
The expression within the square root of (\ref{eigs}) is strictly non-negative if we require
that $\lambda\in[0..1]$. The magnitude of the eigenvalues are
\begin{equation}
||\Lambda_\pm|| = \bar{\epsilon}.
\end{equation}
Hence, for $\bar{\epsilon}\le 1$ the numerical scheme is stable; in fact, without 
dissipation ($\bar{\epsilon}=1$)
the scheme is inherently stable and non-dissipative.

\end{document}